# Spreadsheet Hell

Simon Murphy
Codematic Limited
Carlisle
Cumbria
UK
Simon.murphy@codematic.net

**ABSTRACT**

*This management paper looks at the real world issues faced by practitioners managing spreadsheets through the production phase of their life cycle. It draws on the commercial experience of several developers working with large corporations, either as employees or consultants or contractors. It provides commercial examples of some of the practicalities involved with spreadsheet use around the enterprise.*

## 1     ACKNOWLEDGEMENTS

The author would like to thank all the contributors from the smurfonspreadsheets blog for sharing their commercial experiences.

## 2     INTRODUCTION

60% of large companies feel 'Spreadsheet Hell' describes their reliance on spreadsheets either completely or fairly well. The same survey noted spreadsheet use at 100% of all respondents, the only universal technology. (Durfee, 2004).

It's pretty hard to overstate the importance of spreadsheets to modern business life. (Croll, 2005) found the City of London to be heavily dependent, with most respondents suggesting that spreadsheets were critical to the ongoing viability of the markets and by extension of the City itself.

So we have a completely business critical resource, perhaps like the corporate network or email, and yet in general there appears to be no identifiable person or body responsible for managing it. In many organisations the responsibility falls through the gap between the IT department and the business users. Or it did, Sarbanes Oxley raised the profile of what was once every organisations dirty little secret.

## 3     BACKGROUND

Here are some approximate timings of recent representative spreadsheet based projects undertaken by the contributors:
1. Development = 3 months, live so far 14 months
2. Development = 6 months, live so far 4 years
3. Development = 12 months, live so far 9 years
4. Development = 5 months, live so far 7 years

In all cases the development phase was less than 25% of the total live to date figure, in some cases it is less than 10%. And yet much of the documentation seems to focus on development rather than maintenance. There are very limited published resources to assist in effectively managing up to 90% of the life cycle.

In other business areas the balance could be completely different of course, in particular there are a significant proportion of single use models in some fields.

'Spreadsheet hell' is at least as much about poor management as it is poor quality development. In fact many practitioners feel the mismanagement of production spreadsheets to be the single most significant risk factor in using spreadsheets. One contributor coined the phrase 'versionitis' to describe the uncontrolled proliferation of spreadsheets that seems to occur under poor management conditions.

During a recent spreadsheet risk awareness training session carried out by the author:
- 60% of the delegates felt their Excel skills were inadequate for their job.
- 60% had less than the equivalent of 2 days Excel training.

A list of possible training options and work changes was rated, the top two options were:
- 45% rated expert desk side support as the most useful training option
- 36% felt that a little extra time to deliver would be the most useful change, and in many cases more useful than additional training.

This small group session can not be considered statistically relevant, but the findings do support anecdotal evidence. Any serious program to manage the spreadsheet resource would need to address this perceived time and skill shortfall.

## 4    SPREADSHEET HELL

The spreadsheet issue can be considered on two levels, micro and macro. Spreadsheet hell at the micro level refers to 'frankensheets' (Bruce, 2006). These are big, ugly spreadsheet monsters that are hard to understand, hard to use and hard to test. At the macro level, regardless of the quality (or lack of) of individual spreadsheets the way those spreadsheets are used, shared and replicated creates a whole other level of spreadsheet hell.

This paper covers both, but with more emphasis on the latter. Where possible the points raised are illustrated with a real world example from the contributors' commercial experience.

### 4.1    Micro Level Spreadsheet Hell

An individual spreadsheet can earn itself the frankensheet title fairly easily. Indeed some spreadsheet builders have created nothing else for years. It would be easy to suggest an element of designed in job security were it not for the clear pain the original author experiences when trying to understand their own previous work.

At one company the whole monthly management reporting for 120 business units was driven by an Excel macro one of the consolidation team had recorded/cobbled together. This was a closely guarded treasure and outside interference was not welcome, even though this imposed a significant burden on its owner each period end, especially as it failed most months. As this was an 8 hour process that ran overnight, any failure meant all financial reporting was delayed at least a day. Often the total delay, during which the business had no knowledge of its recent performance was 2-3 days. Although rooted in the individual spreadsheet, this key man dependency has implications at the macro level too.

Several contributors explicitly mentioned this secretiveness and the inevitable key man dependency as key problems (and a common problem), not least because it almost guarantees poor or non existent documentation. "Don't touch my baby" syndrome was how one contributor described it.

Any spreadsheet that is difficult to demonstrate as fit for purpose automatically contributes to spreadsheet hell. And many spreadsheets would need re-writing as part of that assurance process. In reality building a spreadsheet twice still probably leaves it cheaper and quicker than most of the alternatives, but is still pretty rare. In one regulatory reporting project the spreadsheet version cost £30k versus an estimated £500k in a specialised product, it could have been re-written over 10 times, and still been cheaper.

More common than multiple builds is to release the first version to business users after a cursory review and hope the users will spot and report any glaring errors. E.g. one data gathering template was issued with several important balance sheet codes missing, soon fixed once highlighted.

There is a mixed level of agreement on what represents best practice at the individual spreadsheet cell level. E.g. some people think range names are extremely valuable, others don't. However most practitioners would agree on the basic aims of best practice, as being to make understanding and testing reasonably straightforward.

Many factors can affect understandability of a spreadsheet, and some of these have been covered in some depth in previous Eusprig papers. Most practitioners will have favourite techniques, and features they avoid, usually based on good or bad personal experiences. This paper makes no attempt to recommend any particular approach over another, instead it focuses on the wider management story.

**4.2   Macro Level Spreadsheet Hell**

Many organisations are now producing corporate spreadsheet development guidelines, and that is very worthwhile. However, very few seem to have invested in technical infrastructure, either to support the development or the production phase. For example very few spreadsheet developers use the development edition of Office, with integrated source control, or any other form of source control.

More and more organisations are using Office admin policies to restrict access to certain features, for example a recent client blocks all access to the Visual Basic for Applications editor. It is freely available on request, but not by default, thus allowing the organisation to control how certain features are used and by whom. Contrast with the default installation of Office 2000 that many organisations deployed, that had full access to VBA, but no VBA help. No wonder so many people got in a mess.

Ideally a program of spreadsheet management would include
- policies on when to use, and when not to use, spreadsheets
- procedures for safe and effective development of valuable spreadsheets
- features to use and those to avoid, with justifications
- adequate training and coaching appropriate to job role
- procedures and policies for managing the modification of live systems
- policies for safely archiving retired spreadsheets
- Full consideration of all aspects of the systems life cycle

Anecdotal evidence suggests the amount of effort spent keeping a spreadsheet working during its lifetime is inversely related to the quality of the product arriving in the live environment. One rather badly implemented spreadsheet took 3 days to correct, where a well built one would never have gone wrong. At one client, one particular model requires approximately 15 developer days of effort to implement a report headings change, which happens each quarter, another model requires just 1 day to make the same changes.

Keeping these complex spreadsheet systems working is only part of the problem. Another significant issue is extracting and using the business insight locked up in these rigid structures. In one example a stockbroker had 400 workbooks analysing individual stocks. They then needed to summarise which stocks had Price/Earnings ratio below a certain level. Poorly designed spreadsheet systems do not encourage this sort of slice and dice analysis.

**4.3    Spreadsheet use**

There are two broad categories of ongoing spreadsheet use:
1. **Normal use** – for many reporting applications this is the generation of regular reports, also included in here are any year end roll overs.
2. **Changes** – reasons to make non 'normal use' changes to live workbooks will be discussed below

**Normal use**
Normal use can be a significant contributory factor in spreadsheet hell.
Many organisations have limited document management tools, and therefore much essential information must be stored in the file path and name explaining what version the file is and where it is to be used. Many of these files are extremely similar, perhaps 90% of the content is the same reference data or prior period results.

One common file structure is to have a folder for each month and then store the appropriate months spreadsheets (with the same file name) in there each month. In Excel it is not possible to open 2 workbooks with the same name at the same time, even if they are from different folders, so this approach instantly causes reconciliation challenges. The files must be temporarily renamed so they can be compared, failure to reset the names will almost certainly break a linked consolidation report somewhere. As time goes on multiple copies of history proliferate. For example by December there are 12 copies of Januarys results. This makes back posting (changing results of a previously checked and closed period) incredibly easy, and really every months results should be checked each month. Indeed at one clients the February numbers reported in February, March, and April were all (wildly) different, leading to significant consternation amongst the users.

It should be clear that this structure leads to huge duplication, and massive proliferation of very similar spreadsheets, a significant management problem. One very specific risk increased by having many similar spreadsheets floating around is the risk of using the wrong version. This can be exacerbated by a poor or non-existent naming conventions, or unstable file locations. Use of some form of document management system could be a quick win here.

Another common driver of file structure is the use of external links. This feature is widely considered extremely dangerous, but it is such a fast way to build complex systems, it is the norm. In one extreme case a monthly main board pack (approx 20 interlinked workbooks) had to be stored in its own folder each period and never recalculated as it contained an external link circular reference which meant it produced a different number each calculation. The only way to 'lock' in on the published numbers was to archive the reported version somewhere it would not get recalculated before starting the following months reporting.

It is important to remember that spreadsheets are usually the presentation layer of a long chain of data manipulations. Unfortunately this leaves them susceptible to problems introduced by changes anywhere in the chain. One contributor had to hide from an angry manager who had to recall a widely distributed set of reports. The cause was a data supplier unilaterally changing the structure of the data they supplied with no warning. Its no coincidence this example is mentioned in the normal use section, it's a common problem.

**Changes**
There are several basic reasons to make non normal changes to a live workbook.
1. To add a feature
2. To fix an error
3. To improve the design
4. To improve performance
5. To update embedded reference data

In all cases if multiple copies of the spreadsheet are scattered around the network or the world, then coordination and consistency will be a challenge. Very few inexperienced developers make version information obvious enough.

One very real problem with making changes to a live model is the high chance of unexpected side effects. Again, to an extent this is a factor of the spreadsheets underlying quality. Spreadsheets, especially where external links are used, are notorious for breaking changes. Fear of side effects can cause maintainers to contort the existing model rather than make simplifying changes. A simple example is converting a fairly simple formula to a complex array formula for fear of inserting a column for the intermediate calculations. The array approach is recommended regularly, and used regularly even though it is known to degrade performance and maintainability for the future.

VBA adds a whole new dimension to these problems, especially the generally poor quality code seen in many Excel applications. Some code will even overwrite corrections with incorrect values or formulas, often when least expected.

## 5    CONCLUSION

Commercial use of spreadsheets raises issues well beyond the quality of individual models. The overall process of managing the use of this critical resource can have a dramatic effect on the risks to which an organisation is exposed and the value it can leverage from its investments.

## 6    REFERENCES

Croll, G. (2005), "The Importance and Criticality of Spreadsheets in the City of London", Eusprig Proceedings 2005.

Bruce, R. (2006) – Excel-l developers list :http://peach.ease.lsoft.com/

Durfee, D. (2004) - http://www.cfo.com/article.cfm/3014451?f=related accessed 28/02/07

## 7 CONTRIBUTORS

Thanks to R. Bruce, H. Grove, R. McLean, M. Syben, D. Wallentin, and to those who gave their advice and feedback anonymously.